%% file: latfukaya.tex
\documentclass{PoS}

\title{Meson correlators in the $\epsilon$-regime of two-flavor lattice QCD}

\ShortTitle{Meson correlators in the $\epsilon$-regime}

\author{
  JLQCD collaboration:
  \speaker{H.~Fukaya}$^{,a,}$\thanks{E-mail: hfukaya@riken.jp},
  S.~Aoki$^{b,c}$,
  S.~Hashimoto$^{d,e}$,
  T.~Kaneko$^{d,e}$,
  H.~Matsufuru$^d$,
  J.~Noaki$^d$,
  K.~Ogawa$^f$
  T.~Onogi$^g$ and
  N.~Yamada$^{d,e}$
  \vspace*{2mm}
  \\
  \llap{$^a$}
  Theoretical Physics Laboratory, RIKEN, Wako 351-0198, Japan
   \llap{$^b$}
  Graduate School of Pure and Applied Sciences,
  University of Tsukuba, Tsukuba 305-8571, Japan
  \\
  \llap{$^c$}
  Riken BNL Research Center, 
  Brookhaven National Laboratory, Upton, NY 11973, USA
  \\
  \llap{$^d$}
  High Energy Accelerator Research Organization (KEK),
  Tsukuba 305-0801, Japan
  \\
  \llap{$^e$}
  School of High Energy Accelerator Science,
  the Graduate University for Advanced Studies (Sokendai),
  Tsukuba 305-0801, Japan
  \\
  \llap{$^f$}
  Physics Department and Center for Theoretical Sciences, 
National Taiwan University, Taipei, 10617, Taiwan
  \\
  \llap{$^g$}
  Yukawa Institute for Theoretical Physics,
  Kyoto University, Kyoto 606-8502, Japan
}

\abstract{
We calculate the meson correlators 
in the $\epsilon$-regime of two-flavor QCD.
On a $16^3\times 32$ lattice with $a\sim 0.11$~fm,
the lattice simulations are performed with the dynamical overlap fermions.
We reduce the sea quark mass down to $\sim$ 3~MeV and 
the valence quark masses are taken in the range 1--4~MeV.
The meson correlators in various channels are compared with
the predictions of (partially quenched) 
chiral perturbation theory (ChPT).
Including the NLO order of the $\epsilon$-expansion,
we extract the leading-order low energy constants of ChPT, the
pion decay constant $F$ and the chiral condensate $\Sigma$, as
$F=87.3(5.5)$~MeV
and
$\Sigma^{\overline{\mathrm{MS}}}=[237.8(4.0)\mbox{~MeV}]^3$.
}

\FullConference{The XXV International Symposium on Lattice Field Theory\\
		 July 30-4 August 2007\\
		 Regensburg, Germany}

\begin{document}

\input{sec1.tex}
\input{sec2.tex}

\input{sec3.tex}

\input{sec4.tex}


HF thanks P.~H.~Damgaard, T.~DeGrand and P.~Hasenfratz 
for fruitful discussions.
Numerical simulations are performed on Hitachi SR11000 and IBM 
System Blue Gene Solution at High Energy Accelerator Research
Organization (KEK) under a support of its Large Scale Simulation
Program (No. 0716), also in part on
NEC SX-8 at YITP, Kyoto University. 
This work is supported in part by the Grants-in-Aid for
Scientific Research from the Ministry of Education,
Culture, Sports, Science and Technology.
(Nos. 17740171, 18034011, 18340075, 18740167, 18840045, 19540286,
      19740160).

\input{ref.tex}
\end{document}

%% file: sec1.tex
\section{Introduction}

Determination of the leading-order parameters of chiral perturbation
theory (ChPT), {\it i.e.} the chiral condensate $\Sigma$ and the pion decay
constant $F$, is the first goal for lattice QCD in establishing the link between QCD
and ChPT.
This is not an easy task, because the large volume limit must be taken before
taking the chiral limit.
Both limits require enormous computational cost.

Recently, another approach is getting popular, that is the lattice calculation
in the $\epsilon$-regime.
Here, one reduces the quark mass very close to the chiral limit at a fixed
volume $V$. 
Of particular interest is the region where the pion correlation length, or
the inverse pion mass $1/m_\pi$, overshoots the size of the box $L$.
In this regime, the finite volume effect becomes prominent, but it can be
treated in a systematic way within ChPT as an expansion in a new parameter
$\epsilon$ 
\cite{Gasser:1987ah, 
Hansen:1990un, Hasenfratz:1989pk}, 
which has the scale
$m_\pi/\Lambda_{\rm cut} \sim p^2/\Lambda^2_{\rm cut}\sim \epsilon^2$,
where $p$ denotes the pion momentum and 
$\Lambda_{\rm cut}$ is the cutoff of ChPT.
Zero-momentum mode of pion is treated non-perturbatively at the leading order
of the $\epsilon$-expansion.
With this expansion, precise predictions for the volume and topological charge
dependences of the quark condensate, meson correlators, {\it etc}. are
obtained in terms of the low-energy constants in the infinite volume.

Simulating lattice QCD in the $\epsilon$-regime has potential advantages.
Computational cost can be kept manageable by staying in a small box, when
reducing the quark mass until the chiral extrapolation becomes unnecessary.
The remarkable topological charge and quark mass dependences of the meson
correlation functions, for instance, in the $\epsilon$-regime are helpful to
identify the correspondence between QCD and ChPT. 
These nice properties have already been exploited in the quenched QCD studies
\cite{
Bietenholz:2003bj}.
In such works the overlap-Dirac operator \cite{Neuberger:1997fp}
is used, since any tiny violation of chiral symmetry may be amplified in the
$\epsilon$-regime. 

A recent work of the JLQCD collaboration \cite{Fukaya:2007fb}
extended the study in the $\epsilon$-regime to unquenched QCD.
(See also \cite{DeGrand:2006nv} 
for smaller scale works.)
On a $16^3\times32 $ lattice at the lattice spacing 
$a\sim 0.11$~fm (determined from $r_0\sim 0.49$~fm) 
we have generated 460 configurations with two-flavor dynamical overlap quarks 
and Iwasaki gauge action. 
We carried out a simulation at an extremely light sea quark mass $m\sim 3$ MeV,
which is within the $\epsilon$-regime. 
Comparing the Dirac spectrum with the predictions of chiral Random Matrix
Theory, we extracted the value of $\Sigma$ at the leading order of the
$\epsilon$-expansion as
$\Sigma^{\overline{\mathrm{MS}}}(2\mbox{GeV})=[251(7)(11)\mbox{~MeV}]^3$,
where the second error is an estimate of the systematic
error due to the NLO effects of the $\epsilon$-expansion.

In this work, we use the same configurations in the $\epsilon$-regime to
calculate the meson correlators in various channels.
Analytic predictions of ChPT for various channels
are known to the next-to-leading order of the $\epsilon$-expansion
\cite{Damgaard:2001js, Damgaard:2002qe}.
Their extensions to the partially quenched ChPT 
have also become available recently \cite{Damgaard:2007ep}.
We compare these ChPT predictions with the lattice data
to extract $\Sigma$ and $F$ to the NLO accuracy.


%% file: sec2.tex
\section{(Partially quenched) chiral perturbation theory 
in the $\epsilon$-regime at fixed topology}
\label{sec:ChPT}

First, we briefly review the results for the meson correlators 
calculated within (partially quen\\-ched) chiral perturbation theory.
See \cite{Damgaard:2001js, Damgaard:2002qe, Damgaard:2007ep}
for details.
Here we consider $N_v$ valence quarks with a mass $m_v$ and
$N_f=2$ sea quarks with a mass $m_s$, both in the $\epsilon$-regime.

As a fundamental building block of this section, 
let us define the partially quenched
zero-mode partition function which consists of two physical quarks 
(of which mass is $m_s$) and one valence quark (with $m_v$)
and 1 bosonic quark (with $m_b$), at a fixed topology $\nu$.
\begin{eqnarray}
\mathcal{Z}^{\rm PQ}_\nu(\mu_b|\mu_v, \mu_s)
&\equiv& 
\frac{1}{(\mu^2_s-\mu^2_v)^2}
\det \left(
\begin{array}{cccc}
K_\nu(\mu_b) & I_\nu(\mu_v) 
& I_\nu(\mu_s) & I_{\nu-1}(\mu_s)/\mu_s \\
-\mu_b K_{\nu+1}(\mu_b) & \mu_v I_{\nu+1}(\mu_v) 
& \mu_s I_{\nu+1}(\mu_s) & I_{\nu}(\mu_s) \\
\mu^2_b K_{\nu+2}(\mu_b) & \mu^2_v I_{\nu+2}(\mu_v) 
& \mu^2_s I_{\nu+2}(\mu_s) & \mu_sI_{\nu+1}(\mu_s) \\
-
\mu^3_b K_{\nu+3}(\mu_b) & \mu^3_v I_{\nu+3}(\mu_v) 
& \mu^3_s I_{\nu+3}(\mu_s) & \mu^2_sI_{\nu+2}(\mu_s) \\
\end{array}
\right),\nonumber\\
\end{eqnarray}
where $\mu_b=m_b \Sigma V$, $\mu_v=m_v \Sigma V$
and $\mu_s=m_s \Sigma V$. 
$K_\nu$'s and $I_\nu$'s are the modified Bessel functions.
Note that in the limit $\mu_b \to \mu_v$, 
it reproduces the $N_f=2$ full theory partition function
\begin{eqnarray}
\mathcal{Z}^{\rm PQ}_\nu(\mu_v|\mu_v, \mu_s)=
\mathcal{Z}^{\rm full}_\nu(\mu_s)\equiv \det
\left(
\begin{array}{cc}
 I_\nu(\mu_s) & I_{\nu-1}(\mu_s)/\mu_s \\
 \mu_s I_{\nu+1}(\mu_s) & I_{\nu}(\mu_s) 
\end{array}
\right).
\end{eqnarray}
The partially quenched chiral condensate at finite $V$ and $\nu$ is defined by
\begin{eqnarray}
\frac{\Sigma_{\nu}^{{\rm PQ}}(\mu_v,\mu_s)}{\Sigma}
&\equiv& 
-\lim_{\mu_b \to \mu_v}
\frac{\partial}{\partial \mu_b}
\ln \mathcal{Z}^{\rm PQ}_\nu(\mu_b|\mu_v, \mu_s).
\end{eqnarray}
It is not difficult to see that in the $\mu_v \to \mu_s$ limit, 
the partially quenched condensate becomes the one in the full theory,
$\frac{\Sigma_{\nu}^{{\rm PQ}}(\mu_s,\mu_s)}{\Sigma}
=  \frac{\Sigma_{\nu}^{{\rm full}}(\mu_s)}{\Sigma}
\equiv 
\frac{1}{2}\frac{\partial}{\partial \mu_s}\ln 
\mathcal{Z}^{\rm full}_\nu(\mu_s).$
In the following, we also use the double derivative 
of the condensate, defined by
\begin{eqnarray}
\frac{\Delta \Sigma_{\nu}^{{\rm PQ}}(\mu_v,\mu_s)}{\Sigma}
&\equiv& 
\frac{\lim_{\mu_b \to \mu_v}
\partial_{\mu_b} \partial_{\mu_v} 
\mathcal{Z}^{\rm PQ}_\nu(\mu_b|\mu_v, \mu_s)}
{\mathcal{Z}^{\rm full}_\nu(\mu_s)}.
\end{eqnarray}

First, we present the correlation functions of the 
flavored pseudo-scalar and scalar operators,
$
P^a(x)=\bar{q}(x)\tau^a \gamma_5 q(x)$ and
$
S^a(x)=\bar{q}(x)\tau^a q(x),
$
where $\tau^a$ denotes the generator of $SU(N_v)$ group
which the valence quark field $q(x)$ belongs to.
For these correlators, the partially quenched expression 
at fixed topology is known
to ${\cal O}(\epsilon^2)$ (no sum over $a$)
\cite{Damgaard:2007ep}:
\begin{eqnarray}
\label{eq:Pmult}
C_P(t)&\equiv&
\int d^3 x \langle P^a(x)P^a(0)\rangle 
=\frac{1}{2}
\frac{L^3\Sigma_{\rm eff}^2}{\mu_v^{\rm eff}}
\frac{\Sigma_{\nu}^{{\rm PQ}}(\mu_v^{\rm eff}, \mu_s^{\rm eff})}{\Sigma}
-\frac{1}{2}\left[
\frac{2\Sigma^2}{F^2}
\frac{\Delta\Sigma_{\nu}^{{\rm PQ}}(\mu_v,\mu_s)}{\Sigma}
\right.
\nonumber\\
&&
\left.
+\frac{\Sigma^2}{F^2}\frac{\partial_{\mu_v}
\Sigma_{\nu}^{{\rm PQ}}(\mu_v, \mu_s)}{\Sigma}
-\frac{\Sigma^2}{F^2}
\frac{4}{\mu_v^2-\mu_s^2}\left(
\frac{\mu_v\Sigma_{\nu}^{{\rm PQ}}(\mu_v, \mu_s)}{\Sigma}
-\frac{\mu_s\Sigma_{\nu}^{{\rm full}}(\mu_s)}{\Sigma}
\right)
\right]h_1(t/T)
,\\
\label{eq:Smult}
C_S(t)&\equiv&\int d^3 x \langle S^a(x)S^a(0)\rangle 
=
\frac{L^3\Sigma_{\rm eff}^2}{2}
\frac{\partial_{\mu_v}
\Sigma_{\nu}^{{\rm PQ}}(\mu_v^{\rm eff}, \mu_s^{\rm eff})}{\Sigma}
-\frac{1}{2}\left[
\frac{2\Sigma^2}{F^2}
\frac{\nu^2}{\mu_v^2}
\right.
\nonumber\\
&&
\left.
+\frac{\Sigma^2}{F^2}\frac{1}{\mu_v}\frac{
\Sigma_{\nu}^{{\rm PQ}}(\mu_v, \mu_s)}{\Sigma}
-\frac{\Sigma^2}{F^2}
\frac{4}{\mu_v^2-\mu_s^2}\left(
\frac{\mu_v\Sigma_{\nu}^{{\rm PQ}}(\mu_v, \mu_s)}{\Sigma}
-\frac{\mu_s\Sigma_{\nu}^{{\rm full}}(\mu_s)}{\Sigma}
\right)
\right]h_1(t/T)
,
\end{eqnarray}
where $\mu_i^{\rm eff}= m_i\Sigma_{\rm eff}V$, and 
the $t$ dependence of correlators is represented by a function
$h_1(t/T) \equiv 
T\left[\left(t/T-1/2\right)^2-1/12\right]/2$.
$\Sigma$ receives one-loop correction as
$\Sigma_{\rm eff}=\Sigma 
\left(1+3\beta_1/2F^2V^{1/2}\right),$
where $\beta_1$ is the so-called shape coefficient. 
In our numerical study, $\beta_1=0.0836$.

Next, consider the flavored axial-vector and vector operators 
$A^a_0(x)=\bar{q}(x)\tau^a \gamma_0 \gamma_5 
q(x)$ 
and 
$V^a_0(x)=\bar{q}(x)\tau^a \gamma_0 q(x)$.
Their correlators at ${\cal O}(\epsilon^2)$ in the $N_f=2$ ChPT are
(see \cite{Damgaard:2002qe} for details)
\begin{eqnarray}
 C_A(t)\equiv\int d^3x 
 \langle A^a_0(x) A^a_0(0) \rangle 
 & = & -\frac{F^2}{2 T}\left\{{\cal J}^0_+ 
 +\frac{2}{F^2} \left(\frac{\beta_1}{V^{1/2}} {\cal J}^0_+ 
 -\frac{T^2}{V} k_{00} {\cal J}^0_- \right) \right.\nonumber\\
 & & \hspace*{1.75cm}\left. +\frac{4 \mu_s}{F^2} 
 \frac{\Sigma_{\nu}^{{\rm full}}(\mu_s)}{\Sigma} \frac{T^2}{V}h_1(t/T)
\right\}\;,
 \hspace*{0.5cm}\label{fullaxial} \\
C_V(t)\equiv\int d^3 x 
 \langle V^a_0(x) V^a_0(0) \rangle
 & = & -\frac{F^2}{2 T}\left\{{\cal J}^0_- 
 +\frac{2}{F^2} \left(\frac{\beta_1}{V^{1/2}} {\cal J}^0_- 
 -\frac{T^2}{V} k_{00} {\cal J}^0_+ \right) \right\}\;,\label{fullvector}
\end{eqnarray}
where $k_{00}$ is another numerical factor depending on the shape 
of the box (in our case, $k_{00}=0.08331$).
${\cal J}^0_\pm$ are defined by
\begin{eqnarray}
 {\cal J}^0_{\pm} & \equiv & 
 \frac{1}{3} \left( 3 \mp 1 \pm 2 \left[ \frac{\partial_{\mu_s}
\Sigma^{\rm full}_\nu(\mu_s^{\rm eff})}{\Sigma} 
 + 2 \left(\frac{\Sigma^{\rm full}_\nu(\mu^{\rm eff}_s)}{\Sigma}\right)^2 
 + \frac{1}{\mu^{\rm eff}_s} \frac{\Sigma^{\rm full}
_\nu(\mu^{\rm eff}_s)}{\Sigma} 
 - 2 \frac{\nu^2}{(\mu^{\rm eff}_s)^2} \right]
 \right)\;, 
\end{eqnarray}

Note that $C_P(t)$ and $C_S(t)$ are sensitive to $\Sigma$ and $F$ enters only
at ${\cal O}(\epsilon^2)$.
For $C_A(t)$ and $C_V(t)$, on the other hand, $F$ appears in the leading term
and thus can be extracted efficiently.

%% file: sec3.tex
\section{Lattice simulations}
\label{sec:simulation}

Here we summarize our numerical set up for the
simulations. For other details, see \cite{Fukaya:2007fb}.
Our lattice size is  $16^3\times 32$ and the lattice spacing
is determined as $a$ = 0.1111(24)~fm from the heavy quark potential
assuming $r_0$ = 0.49~fm.
We use the overlap fermion \cite{Neuberger:1997fp},
of which the Dirac operator with a quark mass $m$ is given by
\begin{equation}
\label{eq:ov}
D(m)=\left(m_0 + \frac{m}{2}\right)
+\left(m_0 - \frac{m}{2}\right)\gamma_5 \mbox{sgn}
[H_w(-m_0)],
\end{equation}
where $H_W(-m_0)=\gamma_5D_W(-m_0)$ denotes the standard Hermitian
Wilson-Dirac operator. 
We choose $m_0=1.6$ throughout this work. 
(Here and in the following the mass parameters are given
in the lattice unit, unless otherwise stated.)
For the gauge action, we use the Iwasaki action 
at $\beta=2.35$ together with an additional determinant factor corresponding
to Wilson fermions and associated twisted-mass ghosts 
\cite{
Fukaya:2006vs},
which forbids the topology changes 
along the Monte Carlo updates.
In this work, topological charge is fixed to $\nu=0$.
We use the hybrid Monte Carlo (HMC) algorithm.
The sign function in (\ref{eq:ov}) is approximated
by a rational function with Zolotarev's coefficients
after projecting out a few lowest-lying eigenmodes.


In this work, we focus on the run at the lightest sea quark mass $m=0.002$,
which corresponds to $\sim$ 3~MeV in the physical unit and the system is 
well within the $\epsilon$-regime.
We accumulated 4,600 trajectories after discarding
400 trajectories for thermalization.
At every 10 trajectories, we calculated the meson correlators
in various channels.
We take four values of valence quark mass,
$m$ = 0.0005, 0.001, 0.002, and 0.003 (1--4~MeV).
The inversion of the Dirac operator is performed 
simultaneously for all the valence quark masses 
using the multimass solver.

For the meson correlators, the purely low-mode contribution 
(both quark and anti-quark propagators are represented by the low-lying
eigenmodes) 
is replaced by the low-mode averaged (LMA) one, {\it i.e.}
the source point is averaged over all lattice sites.
We find that LMA with 100 eigenmodes improves the statistical signal
substantially for the pseudo-scalar and scalar correlators
while the improvement is marginal for the axial-vector and vector correlators. 

The jackknife bin-size is chosen as 20, with which the statistical error
saturates. 



%% file: sec4.tex
\section{Numerical results} 
\label{sec:results} 

Now we present the numerical results.

For the axial current we use the local operator $A^a_0(x)$,
which receives finite renormalization.
We calculated the renormalization factor using the RI/MOM scheme
and obtained $Z_A$ = 1.3513(13).
In the following, we neglect this tiny statistical error for $Z_A$.

\begin{figure*}[bt]
  \centering
  \includegraphics[width=11cm]{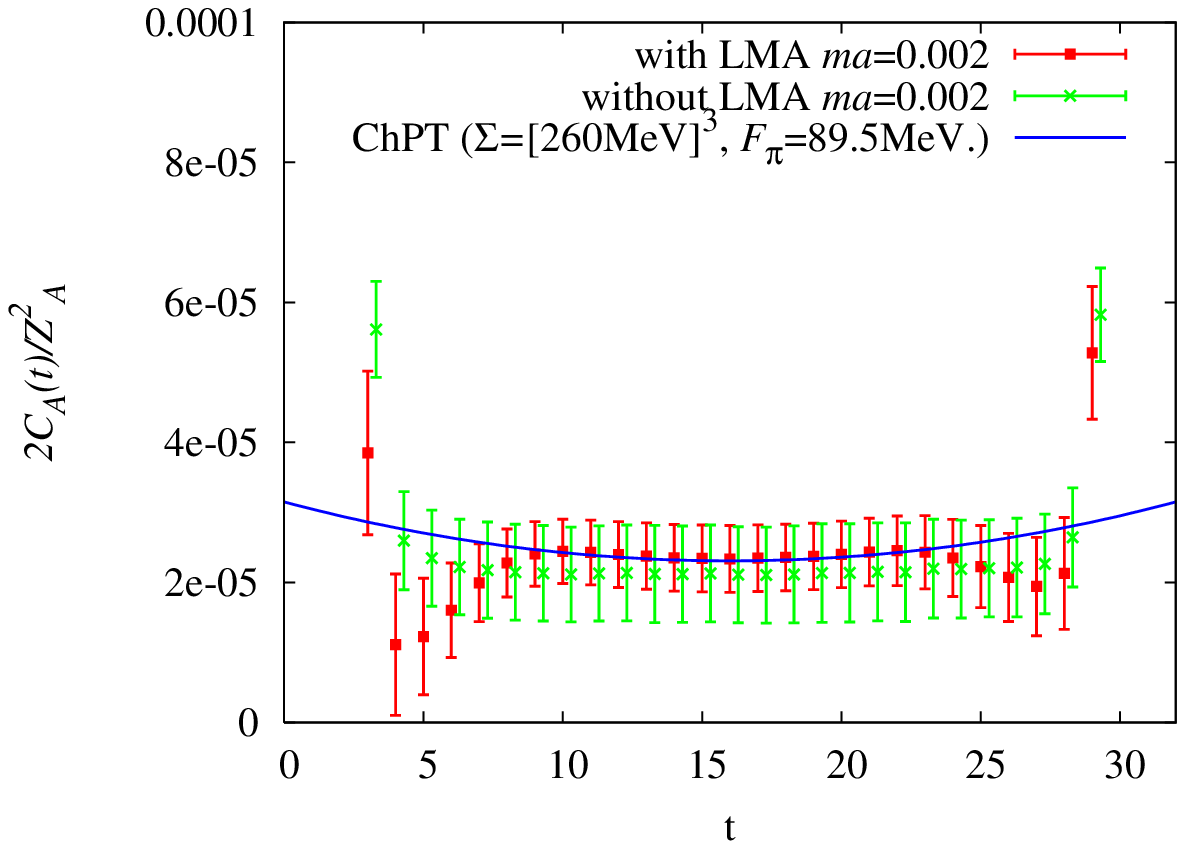}
  \includegraphics[width=11cm]{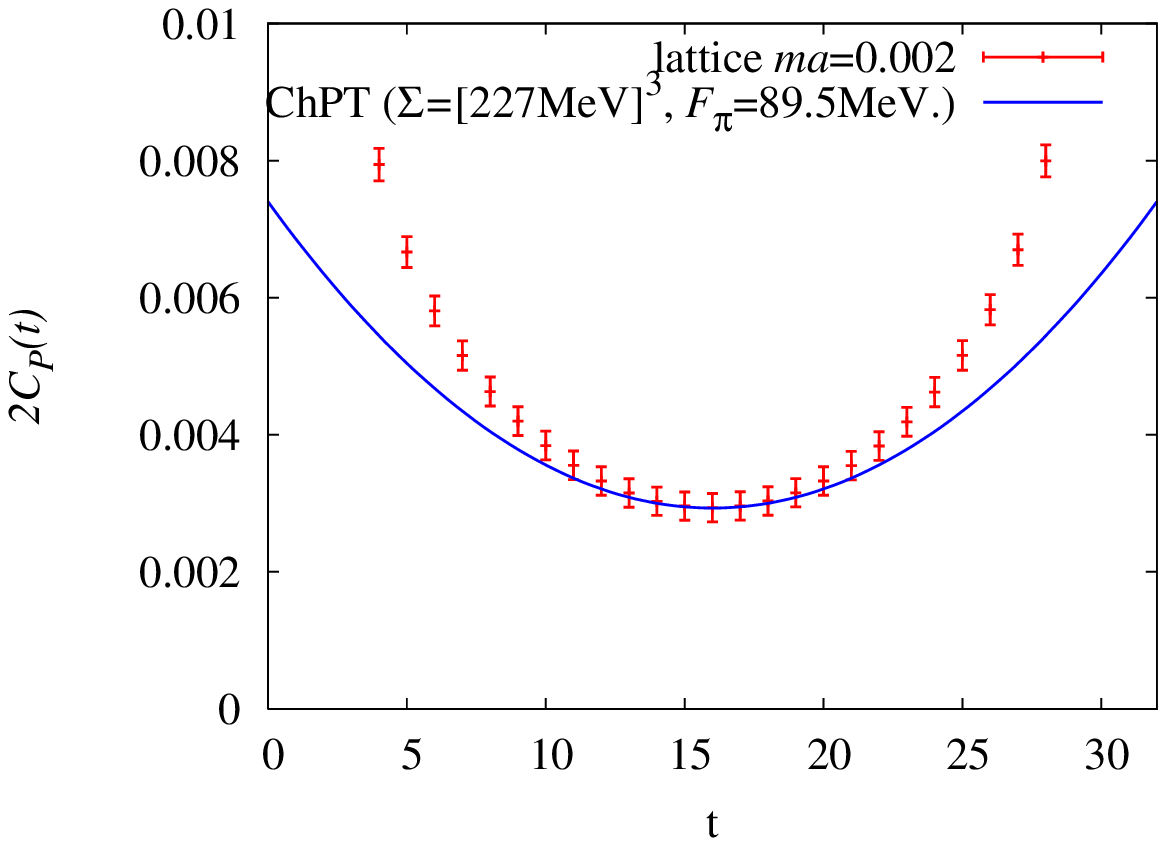}
  \caption{
 Upper figure: the axial vector correlator in the $\epsilon$-regime.
 Filled square symbols denote the low-mode averaged correlator
 while the crosses are not averaged.
 The solid curve shows the best fit yielding
 $\Sigma \sim [260\mbox{~MeV}]^3$ and $F\sim$ 90~MeV).
Lower figure: the pseudo-scalar correlator.
 The solid curve shows the best fit yielding
 $\Sigma\sim [227\mbox{~MeV}]^3$ ($F=89.5$ MeV is given as an input).
  }
  \label{fig:axial}
\end{figure*}

Fig.~\ref{fig:axial} (left panel) shows the data for the axial-vector
correlator at $m_v=m_s=0.002$. 
A two-parameter fit to (\ref{fullaxial}) works well 
($\chi^2/\mbox{d.o.f}\sim 0.01$)
for the low-mode averaged data (filled squares) as shown by a blue curve.
The fitting range is $t\in [12, 20]$.
From this fit we obtain $\Sigma \sim [260(32)\mathrm{~MeV}]^3$ and 
$F\sim 90(6)$ MeV. 
The statistical error in $\Sigma$ is large ($\sim$ 30\%) because $\Sigma$
appears only at ${\cal O}(\epsilon^2)$, while $F$ is determined to a good
precision.

Next, let us look at the pseudo-scalar channel at $m_v=m_s=0.002$.
Using $F$ obtained via the axial-vector correlator as an input,
we fit the pseudo-scalar channel with (\ref{eq:Pmult}) and 
obtain $\Sigma$ with much better precision.
Fig.~\ref{fig:axial} (right panel) shows the fit curve with a 
fit range $t\in [12,20]$ ($\chi^2/\mbox{d.o.f}=0.07$).
A two-parameter fit with $F$ as another free parameter does not work because
the sensitivity to $F$ is too weak. 

To obtain the best result we perform a simultaneous fit to both $P^aP^a$ and
$A^a_0A^a_0$ correlators and obtain
$\Sigma = [227.6(3.7)\mbox{~MeV}]^3$ and
$F$ = 87.3(5.6)~MeV (or $\sqrt{2}F$ = 123.5(7.9)~MeV),
where the statistical error in $a$ is also taken into account.
Here the fit range is $t\in [12,20]$ and $\chi^2/\mbox{d.o.f}=0.02$.
Multiplying the renormalization factor calculated non-perturbatively through
the RI/MOM scheme, 
we obtained the renormalized condensate 
$\Sigma^{\overline{\mathrm{MS}}}
(\mathrm{2~GeV}) = [239.8(4.0)\mbox{~MeV}]^3$.
This value is consistent
with our previous result obtained
through the Dirac spectrum
$\Sigma^{\rm Dirac}=[251(7)(11)\mbox{~MeV}]^3$.



\begin{figure*}[t]
  \centering
  \includegraphics[width=7.5cm]{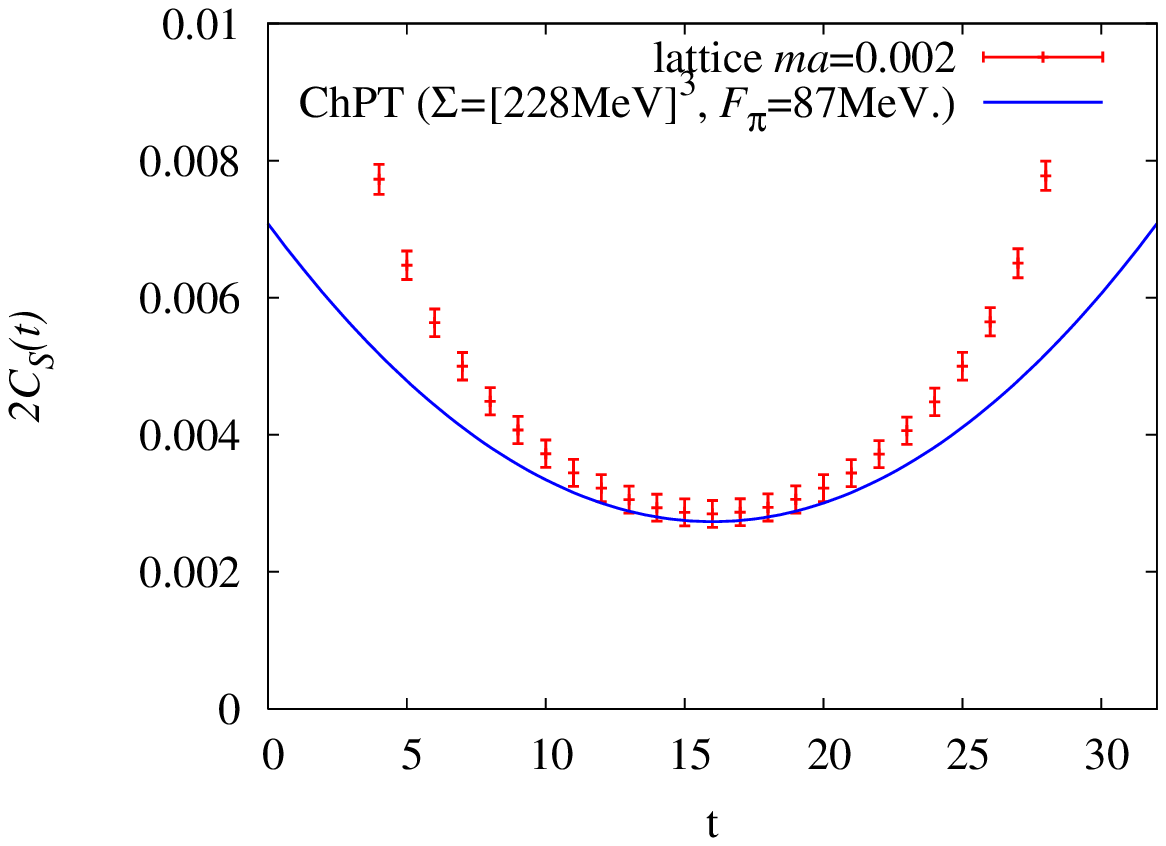}
  \includegraphics[width=7.5cm]{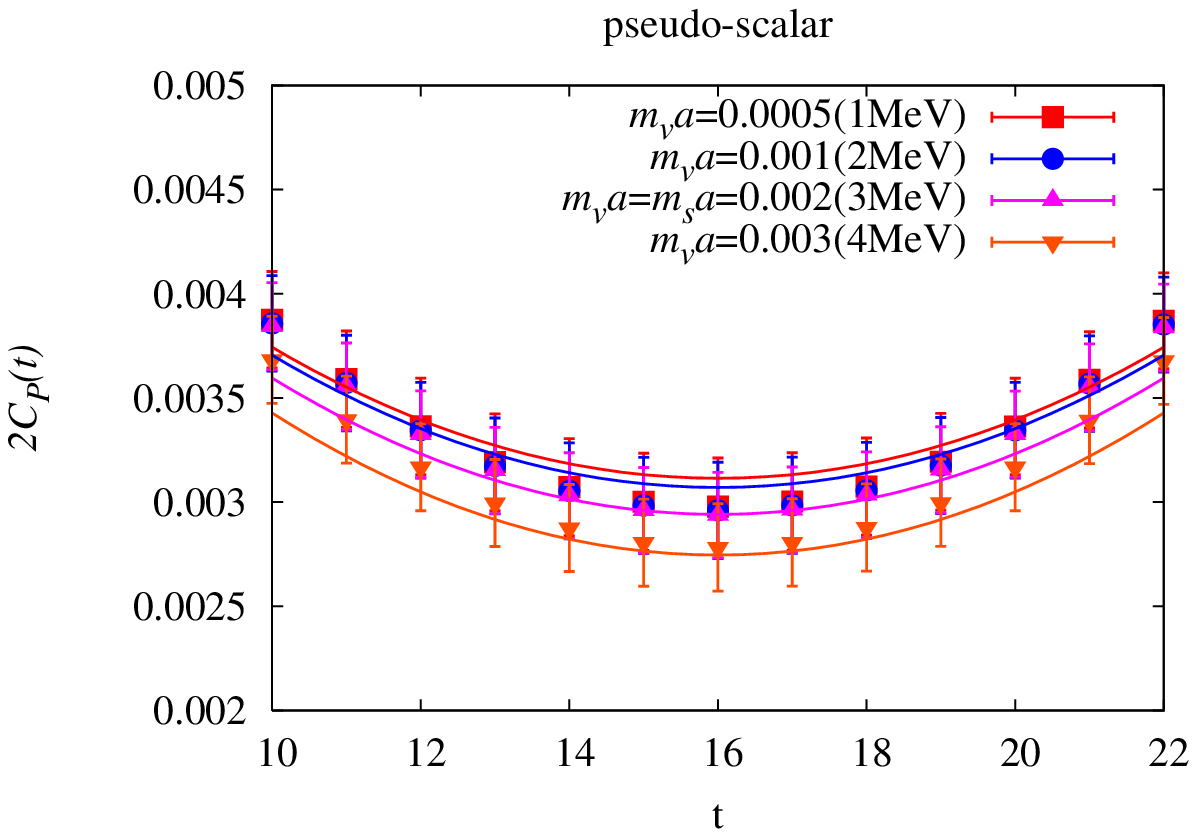}
  \caption{
 The scalar (left) and the partially quenched pseudoscalar 
(right) correlators 
 in the $\epsilon$-regime.
 The solid curves represent the ChPT prediction with
$\Sigma = [227.6\mbox{MeV}]^3$ and
$F = 87.3\mbox{MeV}$ (No free parameter left).
  }
  \label{fig:scalar}
\end{figure*}

Once the parameters $\Sigma$ and $F$ are determined, there is no additional
free parameters at the given order of the $\epsilon$-expansion, and therefore
the comparison provides a stringent test of the lattice data and/or the
$\epsilon$-expansion. 
For instance, the lattice data for the scalar channel is shown in
Fig.~\ref{fig:scalar} (left panel).
The curve in the plot is {\it not} a fit to the data but a ChPT prediction
(\ref{eq:Smult}).
The agreement is remarkable.
Furthermore, we also test the consistency with the partially quenched data
sets ($m_v\not= m_s$) for the pseudo-scalar channel (Fig.~\ref{fig:scalar}
(right panel)).
The curves showing the ChPT prediction without free parameters are
perfectly consistent with the lattice data.

\section{Discussions}
We have compared the different channels
(pseudo-scalar, scalar, axial-vector and vector), 
partially quenched correlators with four different valence quark masses,
and Dirac spectrum, and they are all consistent with each other.
For the change of the fitting range within $t_{\rm min}\in [10, 15]$, 
both $\Sigma$ and $F$ are quite stable (within 1\%) with similar error-bars.
For the finite volume correction, the NLO in the
$\epsilon$-expansion, {\it i.e.} ${\cal O}(\epsilon^2)\sim p^2\sim 1/L^2$,
is taken into account within ChPT.

Still, we find some discrepancy in the pion decay constant with
the preliminary result ($F$ = 78(3)(1)~MeV) obtained in the $p$-regime using
the NNLO formula for the chiral extrapolation \cite{Noaki}.
(The $p$-regime simulation has been done at a slightly coarser lattice
spacing, $a\sim$ 0.12~fm, but we do not expect substantial discretization
effect by this small change of $a$.)
This may signal some unknown source of systematic error, possibly the
higher order contributions  in either the $\epsilon$-expansion or the
$p$-expansion.\\

%% file: latfukaya.bbl
\begin{thebibliography}{9}


\bibitem{Gasser:1987ah}
  J.~Gasser and H.~Leutwyler,
  Phys.\ Lett.\  B {\bf 188}, 477 (1987).
\bibitem{Hansen:1990un}
  F.~C.~Hansen,
  Nucl.\ Phys.\  B {\bf 345}, 685 (1990).

\bibitem{Hasenfratz:1989pk}
  P.~Hasenfratz and H.~Leutwyler,
  Nucl.\ Phys.\  B {\bf 343}, 241 (1990).



\bibitem{Bietenholz:2003bj}
  W.~Bietenholz, T.~Chiarappa, K.~Jansen, K.~I.~Nagai and S.~Shcheredin,
  JHEP {\bf 0402}, 023 (2004),
  L.~Giusti, P.~Hernandez, M.~Laine, P.~Weisz and H.~Wittig,
  JHEP {\bf 0401}, 003 (2004),
  L.~Giusti, P.~Hernandez, M.~Laine, P.~Weisz and H.~Wittig,
  JHEP {\bf 0404}, 013 (2004),
  H.~Fukaya, S.~Hashimoto and K.~Ogawa,
  Prog.\ Theor.\ Phys.\  {\bf 114}, 451 (2005).


\bibitem{Neuberger:1997fp}
  H.~Neuberger,
  Phys.\ Lett.\  B {\bf 417}, 141 (1998),
  H.~Neuberger,
  Phys.\ Lett.\  B {\bf 427}, 353 (1998).


\bibitem{Fukaya:2007fb}
  H.~Fukaya {\it et al.}  [JLQCD Collaboration],
  Phys.\ Rev.\ Lett.\  {\bf 98}, 172001 (2007),
  H.~Fukaya {\it et al.},
  Phys.\ Rev.\  D {\bf 76}, 054503 (2007).




\bibitem{DeGrand:2006nv}
  T.~DeGrand, Z.~Liu and S.~Schaefer,
  Phys.\ Rev.\  D {\bf 74}, 094504 (2006)
  [Erratum-ibid.\  D {\bf 74}, 099904 (2006)],
  P.~Hasenfratz, D.~Hierl, V.~Maillart, F.~Niedermayer, A.~Schafer, C.~Weiermann and M.~Weingart,
  arXiv:0707.0071 [hep-lat],
  T.~DeGrand and S.~Schaefer,
  arXiv:0708.1731 [hep-lat].

\bibitem{Fukaya:2006vs}
  H.~Fukaya, S.~Hashimoto, K.~I.~Ishikawa, T.~Kaneko, H.~Matsufuru, T.~Onogi and N.~Yamada
                  [JLQCD Collaboration],
  Phys.\ Rev.\  D {\bf 74}, 094505 (2006).


\bibitem{Damgaard:2001js}
  P.~H.~Damgaard, M.~C.~Diamantini, P.~Hernandez and K.~Jansen,
  Nucl.\ Phys.\  B {\bf 629}, 445 (2002).

\bibitem{Damgaard:2002qe}
  P.~H.~Damgaard, P.~Hernandez, K.~Jansen, M.~Laine and L.~Lellouch,
  Nucl.\ Phys.\  B {\bf 656}, 226 (2003).


\bibitem{Damgaard:2007ep}
  P.~H.~Damgaard and H.~Fukaya,
  arXiv:0707.3740 [hep-lat].

\bibitem{Noaki}
 JLQCD collaboration (J.Noaki {\it et al}.), in these proceedings.

\end{thebibliography}
